\documentclass[11pt]{article}

\usepackage{epsf,latexsym,amssymb}
\usepackage{graphicx}
\usepackage{amsmath}  

\newcommand{\pic}[2]{\includegraphics[scale=#1]{#2}}
\newcommand{\picc}[2]{\begin{center}\pic{#1}{#2}\end{center}}

\newcommand{\comment}[1]{}
\newcommand{\edo}{\end{document}}

\newcommand{\R}{{\mathbb R}}  

\newtheorem{theorem}{Theorem}
\newtheorem{itlemma}{Lemma}[section] 
\newtheorem{itproposition}[itlemma]{Proposition}
\newtheorem{itcorollary}[itlemma]{Corollary}
\newtheorem{itremark}[itlemma]{Remark}
\newtheorem{itdefinition}[itlemma]{Definition}
\newtheorem{itexample}[itlemma]{Example}

\newenvironment{lemma}{\begin{itlemma}\rm}{\end{itlemma}} 
\newenvironment{remark}{\begin{itremark}\rm}{\end{itremark}} 
\newenvironment{corollary}{\begin{itcorollary}\rm}{\end{itcorollary}}
\newenvironment{proposition}{\begin{itproposition}\rm}{\end{itproposition}}
\newenvironment{definition}{\begin{itdefinition}\rm}{\end{itdefinition}}
\newenvironment{example}{\begin{itexample}\rm}{\end{itexample}}

\newcommand{\be}[1]{\begin{equation}\label{#1}}
\newcommand{\ee}{\end{equation}}
\newcommand{\bl}[1]{\begin{lemma}\label{#1}}
\newcommand{\ble}[1]{\begin{lemmaex}\label{#1}}
\newcommand{\br}[1]{\begin{remark}\label{#1}}
\newcommand{\bt}[1]{\begin{theorem}\label{#1}}
\newcommand{\bd}[1]{\begin{definition}\label{#1}}
\newcommand{\bp}[1]{\begin{proposition}\label{#1}}
\newcommand{\bc}[1]{\begin{corollary}\label{#1}}
\newcommand{\bex}[1]{\begin{example}\label{#1}}
\newcommand{\ec}{\mybox\end{corollary}}
\newcommand{\eex}{\mybox\end{example}}
\newcommand{\eem}{\mybox\end{example}}
\newcommand{\el}{\mybox\end{lemma}}
\newcommand{\er}{\mybox\end{remark}}
\newcommand{\et}{\qed\end{theorem}}
\newcommand{\ed}{\mybox\end{definition}}
\newcommand{\ep}{\mybox\end{proposition}}
\newcommand{\epr}{\end{proof}}
\newcommand{\bpr}{\begin{proof}}

\newcommand{\ecs}{\end{corollary}}
\newcommand{\eexs}{\end{example}}
\newcommand{\els}{\end{lemma}}
\newcommand{\ers}{\end{remark}}
\newcommand{\ets}{\end{theorem}}
\newcommand{\eds}{\end{definition}}
\newcommand{\eps}{\end{proposition}}
\newcommand{\halmos}{\rule{1ex}{1.4ex}}
\newcommand{\qed}{\hfill \halmos} 
\newcommand{\mybox}{\hfill $\Box$} 
\newcommand{\beq}{\begin{eqnarray}}
\newcommand{\eeq}{\end{eqnarray}}
\newcommand{\beqn}{\begin{eqnarray*}}
\newcommand{\eeqn}{\end{eqnarray*}}
\newcommand{\bi}{\begin{itemize}}
\newcommand{\ei}{\end{itemize}}
\newcommand{\ben}{\begin{enumerate}}
\newcommand{\een}{\end{enumerate}}

\newcommand{\bes}[1]{\begin{subequations}\label{#1}\begin{eqnarray}}
\newcommand{\ees}[1]{\end{eqnarray}\end{subequations}}

\newenvironment{proof}{\noindent {\em Proof}.\ }{\hspace*{\fill}$\halmos$\medskip}

\newcommand{\x}{x}  
\newcommand{\f}{f}  
\newcommand{\U}{{\mathbb U}}
\newcommand{\Y}{{\mathbb Y}}
\newcommand{\X}{{\mathbb X}}
\newcommand{\sarr}{\rightarrow } 
\newcommand{\barx}{{\bar \x}}
\newcommand{\baru}{{\bar u}}
\newcommand{\barv}{{\bar v}}
\newcommand{\seq}{=}
\newcommand{\zo}{x_0}
\newcommand{\la}[1]{{#1}_{\mbox{\sc la}}}

\newcommand{\I}{U} 

\newcommand{\Fig}{Fig.}

\newcommand{\ecoli}{\emph{E.\ coli}}

\newcommand{\fractextnp}[2]{#1/#2}

\newcommand{\s}{s}
\newcommand{\pp}{c}

\title{Remarks on invariance of population distributions\\
for 
systems with equivariant internal dynamics} 
\author{Eduardo D. Sontag\\
Rutgers University}

\begin{document}
\maketitle

\medskip

%

\section{Introduction}

There has been recent interest, particularly in the systems biology literature,
in the study of 
symmetry invariances of responses of dynamical systems. %
The paper~\cite{shoval10} obtained sufficient characterizations of symmetry
invariance using a notion of equivariance, and this characterization was shown
to be necessary as well as sufficient in~\cite{shoval_alon_sontag_2011}. %
Both \cite{shoval10} and~\cite{shoval_alon_sontag_2011}
sketched how to extend the results to motile systems that explore space, so
long as the ``motor dynamics'' depends only on an invariant response.
Specifically, these results predicted that \emph{E.\ coli} bacteria would
produce scale-invariant searches, meaning that distributions of bacteria, even
under non-uniform and time-varying chemoeffector fields, should be invariant
under any rescaling of the input field.
This prediction was subsequently experimentally verified
in~\cite{ShimizuStocker2011}. 
In this note, we 
remark that, for a velocity-jump Markov model, the PDE for
the evolution of densities (or normalized concentrations) in time inherits the
symmetry-invariance property from individual behaviors.  Although not at all
surprising, this provides further theoretical justification for passing from
individual-based models 
to population predictions.

\section{Symmetries and equivariances}


We review the general setup in~\cite{shoval10,shoval_alon_sontag_2011}.
Consider dynamical systems with inputs and outputs
\cite{mct},
\be{eq:sys}
\dot \x = \f(\x,u)\,,\quad\quad 
y = h(\x,u)
\,.
\ee
The functions $\f$, $h$ describe respectively the dynamics and the
read-out map.%
\footnote{The results in~\cite{shoval_alon_sontag_2011} were stated for $h$ not
  directly dependent on $u$, but the theory is the same in the more general
  case of $u$-dependence, as was also remarked there.}
Equation~(\ref{eq:sys}) is shorthand for
\[
\frac{d\x}{dt}(t)
=
\f(\x(t),u(t))
\,,\quad\quad 
y(t)=h(\x(t),u(t))\,.
\]
Here, 
$u=u(t)$ is 
a generally time-dependent 
input (stimulus, excitation) function, $\x(t)$ is an 
$n$-dimensional vector of state variables, and $y(t)$ is the output
(response, reporter) variable.
States, inputs, and outputs are constrained to lie in
particular subsets $\X$, $\U$, and $\Y$ respectively, of
Euclidean spaces $\R^n,\R^m,\R^q$.

We assume that for each piecewise-continuous input
$u:[0,\infty )\sarr \U$, and each initial state $\xi \in \X$, there is a
unique solution $\x:[0,\infty )\sarr \X$ of~(\ref{eq:sys}) with initial
condition $\x(0)\seq\xi $, which we write as 
$
\varphi(t,\xi ,u)
$,
and we denote the corresponding output $y:[0,\infty )\sarr \Y$,
given by $h(\varphi(t,\xi ,u),u(t))$, as
$
\psi (t,\xi ,u)
$.
We also assume that for each constant input $u(t)\equiv \baru$, there is a
unique solution $\barx=\sigma (\baru)$ of the algebraic equation
$\f(\barx,\baru)\seq0$.
Often one also assumes that this steady state is globally asymptotically
stable (GAS): it is Lyapunov stable and globally attracting for
the system when the input is $u(t)\equiv \baru$:
$
\lim_{t\rightarrow \infty }\varphi(t,\xi ,u)
=
\sigma (\baru)
$
for every initial condition $\xi \in \X$.
The GAS property is not required for the results to follow, however.


If $\X$ is an open set, or the closure of an open set, in $\R^n$,
the system~(\ref{eq:sys}) is said to be \emph{analytic} if $\f$ and $h$ are
real-analytic (can be expanded into locally convergent power series around
each point) with respect to $\x$, and \emph{irreducible} if it is accessible
and observable. 

An accessible system is one for which the accessibility rank condition holds:
$\la{{\cal F}}(\zo)=\R^n$ for every $\zo\in \X$, where
$\la{{\cal F}}$ is the accessibility Lie algebra of the system.
Intuitively, this means that no conservation laws restrict motions to proper
submanifolds.
For analytic systems, accessibility is equivalent to the property that the set
of points reachable from any given state $\x$ has a nonempty interior; see a
proof and more details in the textbook~\cite{mct}.
An observable system is one for which
$\psi (t,\zo,u) = \psi (t,\widetilde \zo,u)$ for all $u,t$ implies
$\zo = \widetilde \zo$.
Intuitively, observability means that no pairs of distinct states can give
rise to an identical temporal response to all possible inputs.
For analytic input-affine systems, observability is equivalent to the property
that any distinct two states can be separated by the observation space;
see~\cite{mct}, Remark 6.4.2 for a proof and discussion.
In the context of applications to biomolecular systems, analyticity and
irreducibility are weak techinal assumptions, often satisfied. 

\subsubsection*{Adaptation, invariance, and equivariances}

\bd{def:adapt}
The system~(\ref{eq:sys})
\emph{perfectly adapts to constant inputs}
provided that the steady-state output $h(\sigma (\baru),\baru)$ equals some fixed
$y_0\in \Y$, independently of the particular input value $\baru\in \U$.
\ed

That is, the steady-state output value is independent of the actual value of
the input, provided that the input is a constant (a step function).

Invariance will be defined relative to a set
${\cal P}$ of continuous and onto input transformations $\pi :\U\sarr \U$.
For each input $u(t)$ and $\pi \in {\cal P}$, we abuse notation and denote by ``$\pi u$'' 
(even if $\pi $ is nonlinear) the function of time that equals $\pi (u(t))$ at
time $t$. 
(The continuity assumption is only made in order to ensure that $\pi u$ is a
piecewise continuous function of time if $u$ is.
The ontoness assumption, that is, $\pi \U=\U$, and can be weakened considerably:
it is only used in in the main theorem in order to prove that a system
$\dot x=f(x,\pi u)$, $y=h(x,\pi u)$ is irreducible if the original system is
irreducible, but far less than ontoness is usually required for that.)

An example is \emph{scale invariance}, in which $\U=\R_{>0}$
and ${\cal P}=\{u\mapsto pu, p\geq 0\}$.
(Scale invariance is sometimes called ``fold-change detection'' (FCD), since
the only changes that can be detected in a response are those due to different
fold-changes in inputs.)

\bd{def:invariance}
The system~(\ref{eq:sys}) has \emph{response invariance to symmetries in ${\cal P}$} 
or, for short, is \emph{${\cal P}$-invariant}
if
\be{eq:defFCD}
\psi (t,\sigma (\baru),u)\;=\; \psi (t,\sigma (\pi \baru),\pi u) 
\ee
holds for all $t\geq0$, all inputs $u\seq u(t)$, all
constants $\baru$, and all transformations $\pi \in {\cal P}$.
\ed

Under the assumption that the action of ${\cal P}$ is transitive,
i.e., for any two $\baru,\barv\in \U$, there is some $\pi $ such that $\barv=\pi \baru$,
${\cal P}$-invariance implies perfect adaptation,
because the outputs in~(\ref{eq:defFCD}) must coincide at time zero, and any two
inputs can be mapped to each other. 


\bd{def:equivariance}
Given a system~(\ref{eq:sys}) and a set of input transformations ${\cal P}$,
a parametrized set of differentiable mappings
$\left\{\rho _{\pi }:\X\rightarrow \X\right\}_{\pi \in {\cal P}}$ 
is a \emph{${\cal P}$-equivariance family} provided that, for each $\pi $:
\be{eq:equiv}
\f(\rho _{\pi }(\x),\pi u) = (\rho _{\pi })_*(\x)\f(\x,u)\quad\mbox{and}\quad
h(\rho _{\pi }(\x),\pi u)     = h(\x,u)
\ee
for all $\x\in \X$ and $u\in \U$, where
$(\rho _{\pi })_*$ denotes the Jacobian matrix of $\rho _{\pi }$.
If (\ref{eq:equiv}) holds, the system is said to be $\rho _{\pi }$-equivariant under
the input transformation $\pi $.
\ed

The first part of Equation~(\ref{eq:equiv}) is a 
first order quasilinear partial differential equation on the $n$
components of the vector function $\rho _\pi $, for each $u\in U$, and one may
solve such equations, in principle, using the method of characteristics.
The second part of Equation~(\ref{eq:equiv}) is an additional algebraic
constraint on these components.
Observe that the verification of equivariance does not require
the computation of solutions $\psi (t,\sigma (\pi \baru),\pi u)$.
We omit the subscript $\pi $ when clear from the context.



The main result in~\cite{shoval_alon_sontag_2011} is as follows.

\bt{theo:main}
An analytic and irreducible system is ${\cal P}$-invariant if and only if there
exists a ${\cal P}$-equivariance family. 
\et

\br{rem:extensions}
An interesting consequence of this theorem is that, if ${\cal P}$-invariance holds,
then a stronger property holds as well, namely that
\[
\psi (t,x,u)\;=\; \psi (t,\rho (x),\pi u) 
\]
is valid for all $t\geq0$, all inputs $u$, all transformations $\pi \in {\cal P}$,
and \emph{every initial state $x$} (not necessarily a steady state).
Another interesting fact, which follows from the proof of the theorem, is
as follows.  Suppose that we define a ``weakly invariant'' system as one
for which there exists \emph{some} constant $\baru$ such that
(\ref{eq:defFCD}) holds:
$\psi (t,\sigma (\baru),u)=\psi (t,\sigma (\pi \baru),\pi u)$
for all inputs $u$ and all $t\geq 0$
(instead of asking that this holds for every $\baru$).
Then, ``weak invariance'' implies the existence of an equivariance, and hence
also invariance.
The irreducibility property plays a subtle role in these facts.
\er


\section{Symmetry-invariant steering}


We consider next a motile vehicle or organism which explores a space while
measuring the ``intensity'' of an input cue (such as a chemoeffector or light).
The sensed input at time $t$ and position $r$ is $\I(t,r)$, where $r=r(t)$ is
the current position of the vehicle. 
The current position $r(t)$ is derived from the output $y(t)$ of a 
system~(\ref{eq:sys}), through a computation that takes into account the
dynamics of the motor and steering mechanisms.

Deterministic models for such mechanisms are sometimes
appropriate, and one was described in~\cite{shoval10,shoval_alon_sontag_2011}.
An easy argument for that deterministic model shows that, if $y$ is invariant
under symmetries in inputs, then positions $r(t)$ will be invariant under 
symmetry transformations on $\I$.

It is often the case that a more accurate description is one in which the output
$y(t)$ drives a stochastic, not a deterministic, steering mechanism: the
subsystem producing the location $r(t)$ is subject to randomness.

An important instance of this is bacterial \emph{E.\ coli} chemotaxis, where
$y(t)$ represents a signal, the level of phosphorylated protein CheY, which
serves to bias the random switches between tumbling and swimming (``run'')
modes.
Specifically, let us consider the Tu-Shimizu-Berg {\ecoli} chemotaxis model
\cite{TuShimizuBerg2008}, which may be formulated, for realistic parameters
and input levels, as follows:
$\dot m = F_0(y)$, $y=h(m,u)= G(u/e^{\alpha m})$,
where $F_0$ is a decreasing function which crosses zero at some value $y=y_0$
(and $G$ is a suitable function whose precise form is immaterial for
establishing symmetry).
Letting $x:=e^{\alpha m}$ and $F=\alpha F$, we may transform this system into a
``nonlinear integral feedback'' form,
\beqn
\dot x &=& x\, F(h(x,u))\\
h(x,u) &=& G\left(\fractextnp{u}{x}\right)\,.
\eeqn
For this system, homogeneity of $f(x,u)=xF(h(x,u))$ implies scale invariance,
since the unique solution of the equivariance PDE
is 
$\rho (x)=px$, for the scaling symmetry $u\mapsto pu$.
Based on this verification of scale-invariant behavior, \cite{shoval10}
predicted the invariance of distributions of bacteria locations under scalings
of chemoattractant fields.
This prediction was subsequently verified experimentally
in~\cite{ShimizuStocker2011} by means of molecular level analysis of
intracellular signaling (FRET experiments) as well as measurements of swimming
behavior at the level of individual cells and populations (in microfluidic
environments).

\begin{figure}[ht]
\centering
\picc{0.5}{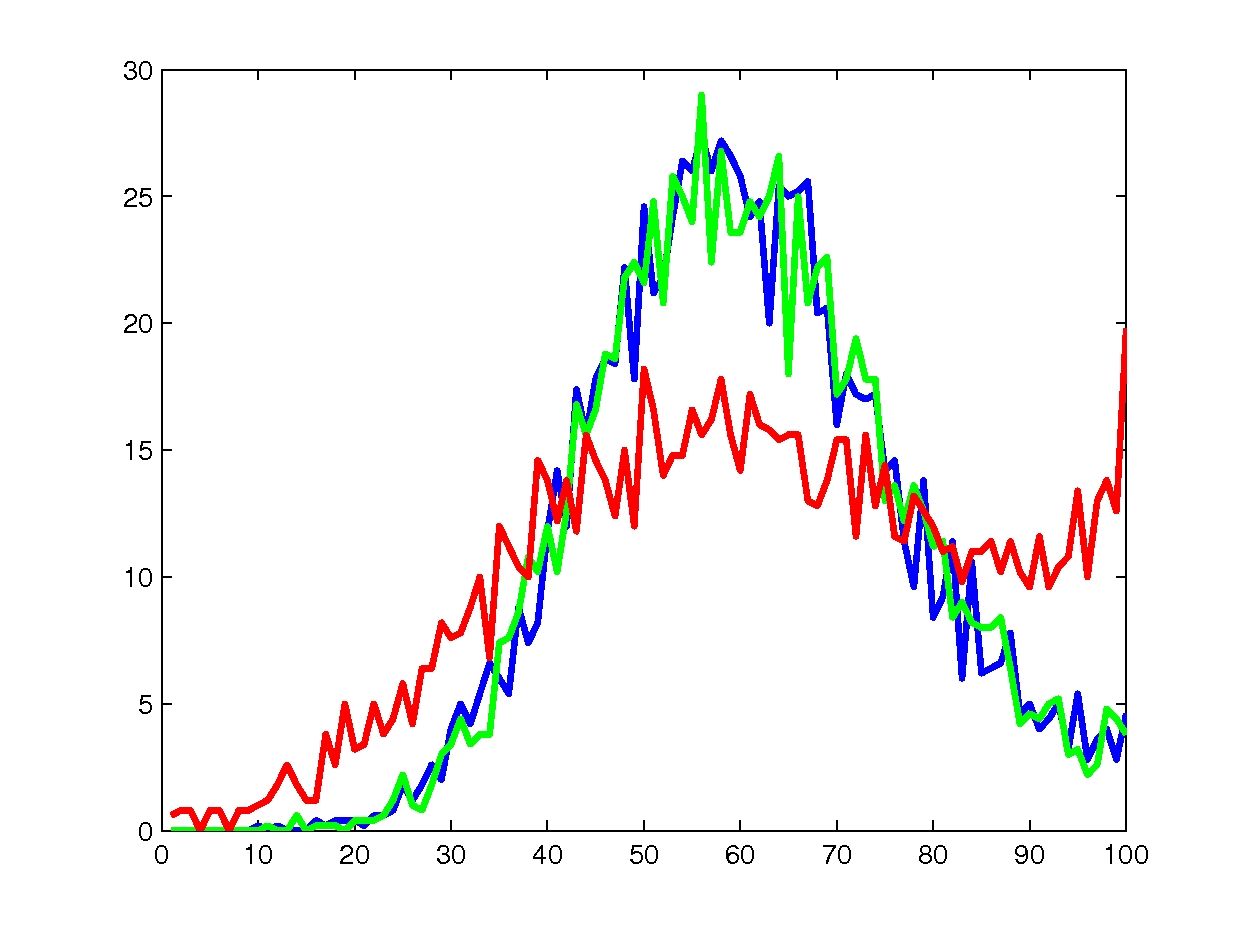}
\caption{Simulations using SPECS code}
\label{fig:specs}
\end{figure}

A simple numerical simulation %
 serves to illustrate the point.
This simulation uses (with no change in parameters) the SPECS agent-based
model for \emph{E.\ coli} chemotaxis that was developed in~\cite{JiangEtAl2010}.%
\footnote{We thank Y.\ Tu for making this code available.}
In this simulation, cells are allowed to swim in a
rectangular channel that is 2000$\mu  m$ long and 400$\mu  m$ wide, and data
is collected in bins of size 20 (so, there are 100 bins along the long axis).
The ligand gradient is stationary and linear (see below for boundary values)
along the length and constant along the width.
We simulated 1000 cells, all initially placed at the middle 
(at length 1000, i.e. bin 50), and plotted the marginal distributions
(along the long axis on which the chemoattractant varies).
Since there behavior is random, the averages of several (five) trials under
each of the conditions are shown.  These average
histograms are plotted for the cell distribution at time $t=500$.
The blue and green histograms in \Fig~\ref{fig:specs} represent, respectively,
results for cells
pre-adapted to a concentration 250 (units are $\mu M$), and linear gradient
$200\ldots 300$, and cells
pre-adapted to a concentration 375, and linear gradient $300\ldots 450$
(a scale change by $p=1.5$).
As expected, the distributions are very similar.
As a control, we also plotted the results of using, once again,
a linear gradient $300\ldots 450$, but now pre-adapting cells to
a concentration of 250.
Since the initial state is not matched, there is no reason for invariance.
Indeed, the resulting red histogram is very different from the previous ones.

One may mathematically formalize probabilistic behavior, and show
symmetry-invariance of search under randomness, in several possible ways.
For instance, in~\cite{shoval_alon_sontag_2011} a simple result was presented on
symmetry-invariance search based on pathwise equality of stochastic processes.
We describe next a different approach, %
that employs the formalism of velocity-jump processes
\cite{OthmerDunbarAlt1988}
with added internal dynamics~\cite{ErbanOthmer2004}.

We wish to model motions in a space $\R^N$ (typically $N=1,2,3$; and we assume
for simplicity that motion can occur on the entire space) of individuals
(bacteria, vehicles, etc) whose internal dynamics are described by the states
$x$ in~(\ref{eq:sys}) and which change velocities as a function of the output
$y$. 
To avoid confusion with the variable $x$ used for the internal state,
we use the letter ``$\s$'' to denote points in the space $\R^N$ in which
movement occurs.
The input $u=u(t,\s)$ represents an external signal
present at time $t$ in location $\s$.
The subset $V\subseteq \R^N$ denotes the space of possible velocities.

We assume that the system can instantaneously change orientations.  (For
{\ecoli} bacteria this would mean that we are ignoring tumble durations.)

The concentration at time $t$ of individuals present at time $t$ in location
$\s$ and having internal state $x$ and velocity $v$ is denoted by
$\pp(t,\s,v,x)$.  
We interpret $\pp(t,\s,v,x)\,d\s\,dv\,dx$ as the number of individuals
located between $\s$ and $\s+d\s$, having velocity between $v$ and $v+dv$,
and whose internal state is between $x$ and $x+dx$. 
Normalized by the total number of individuals, one may also think of $\pp$ as a
probability density, at each time $t$. 

We assume that velocities change at random.  The \emph{times} at which
velocities jump are controlled by a Poisson process with intensity $\lambda (y)$.
Given that a jump in velocity occurs, \emph{which} particular new velocity is
picked is itself the result of a random choice; the kernel $T_y(v,v',y)$ gives
the probability of a change in velocity from $v'$ to $v$.  Since $T$ is a
probability density, $\int_V T_y(v,v')\,dv=1$ for every $y$.  Notice that,
just as with the jump instants, the kernel also depends on the state only
through the output $y$.

Then the evolution (transport, Fokker-Planck, or forward Kolmogorov) equation
for $\pp=\pp(t,\s,v,x)$ is:
\be{eq:Fokker-Planck-Kolmogorov}
\frac{\partial \pp}{\partial t}
+
\nabla_{\s}\cdot \pp v
+
\nabla_{x}\cdot \pp f
\;=\;
-\lambda (y)\pp
+
\int_V
\lambda (y)T_y(v,v')\pp(t,\s,v',x)\,dv'
\ee
where most arguments have been omitted for simplicity, but understood as
holding for all $(t,\s,v,x)$.
(More generally, the right-hand side could be replaced by a more complicated
discrete rate of change, if jumps are governed by a non-Poisson process.)
The input at location $\s$ and time $t$ is $U(t,s)$, and it appears in these
equations through the vector field $f$ in~(\ref{eq:sys}).

Sometimes it is useful to view~(\ref{eq:Fokker-Planck-Kolmogorov})
as a set of partial differential equations indexed and coupled by the
velocities $v$.
For example, when $N=1$ and there is a constant speed $v_0>0$,
$V=\{-v_0,v_0\}$ is a two-element set which provides the orientation of
movement, $T_y(v,v')=1$ (there is only one possible jump, namely a reversal
of direction), and~(\ref{eq:Fokker-Planck-Kolmogorov}) describes a
telegraph-type process: denoting $\pp^+(t,\s,x)=\pp(t,\s,v_0,x)$
and $\pp^-(t,\s,x)=\pp(t,\s,-v_0,x)$, (\ref{eq:Fokker-Planck-Kolmogorov}) can
then be thought of as set of coupled partial differential equations, one for 
$\fractextnp{\partial \pp^+}{\partial t}$ and one for
$\fractextnp{\partial \pp^-}{\partial t}$:
\beqn
\frac{\partial \pp^+}{\partial t}
+
v_0 \frac{\partial \pp^+}{\partial x}
+
\nabla_{x}\cdot f\pp^+
&=& 
\lambda (y)[-\pp^++\pp^-]\\
\frac{\partial \pp^-}{\partial t}
-
v_0 \frac{\partial \pp^-}{\partial x}
+
\nabla_{x}\cdot f\pp^-
&=& 
\lambda (y)[\pp^+-\pp^-]\,.
\eeqn
The reference~\cite{ErbanOthmer2004} discusses mathematical aspects of
the PDE~(\ref{eq:Fokker-Planck-Kolmogorov}), which will not be discussed
here.  We focus, purely formally, on symmetry invariance.

Let us assume given $\pi $ and an associated equivariance $\rho =\rho _\pi $, so that
(\ref{eq:equiv}) holds:
\[
\f(\rho (\x),\pi u) = \rho _*(\x)\f(\x,u)\quad\mbox{and}\quad
h(\rho (\x),\pi u)     = h(\x,u)
\]
for all $\x\in \X$ and $u\in \U$, where
$\rho _*$ denotes the Jacobian matrix of $\rho $.
We will also make the following assumption on the divergence of $f$:
\be{eq:divergence}
(\nabla_{x}\cdot f) (\rho (x),\pi u) \;=\; (\nabla_{x}\cdot f) (x,u)
\ee
for all $\x\in \X$ and $u\in \U$.
This property is automatically satisfied for most of the examples treated
in~\cite{shoval_alon_sontag_2011}, since in these examples, which are for scale
invariance $\pi u=pu$, $\rho $ is a linear mapping.  In general, if $\rho (x)=Rx$ for
a matrix $R$, then the equivariance condition $\f(Rx,\pi u) = R\f(\x,u)$
implies, taking Jacobians, that $f_*(Rx,\pi u) = Rf_*(x,u)R^{-1}$.  Since
two similar matrices have the same trace, and $\nabla_{x}\cdot f$ is the trace of
the Jacobian of $f$, it follows that~(\ref{eq:divergence}) is valid.

Our main observation is that the same distribution of individuals will result
if the input field $U$ is replaced by $\pi U$, provided that the internal
states are transformed by $\rho $.  A precise statement is as follows.

\bt{theo:jumpmarkov1}
Suppose that $c$ satisfies~(\ref{eq:Fokker-Planck-Kolmogorov}) with respect to
an input field $U$.  Define
\[
\widetilde \pp(t,\s,v,x) = \pp(t,\s,v,\rho ^{-1}(x))\,.
\]
Then $\widetilde \pp$ satisfies~(\ref{eq:Fokker-Planck-Kolmogorov}) with respect to
the input field $\pi U$.
\ets

\bpr
We start by writing all the arguments in~(\ref{eq:Fokker-Planck-Kolmogorov})
explicitly:
\beqn
&&\hspace{-30pt}\frac{\partial \pp}{\partial t}(t,\s,v,x)
+
(\nabla_{\s}\cdot \Gamma _1)(t,\s,v,x)
+
(\nabla_{x} \cdot \Gamma _2)(t,\s,v,x)\\
&=&
-\lambda (h(x,U(t,\s)))\pp(t,\s,v,x)\\
&+&
\int_V
\lambda (h(x,U(t,\s)))T_{h(x,U(t,\s))}(v,v')\pp(t,\s,v',x)\,dv'
\eeqn
where
\beqn
\Gamma _1(t,\s,v,x) &=& \pp(t,\s,v,x)v\\
\Gamma _2(t,\s,v,x) &=& \pp(t,\s,v,x)f(x,U(t,\s)) \,.
\eeqn
Since this equation must hold for all $x$, it holds also when $\rho ^{-1}(x)$ is
replaced for $x$, in other words it is also true that
\beqn
&&\hspace{-30pt}\frac{\partial \pp}{\partial t}(t,\s,v,\rho ^{-1}(x))
+
(\nabla_{\s}\cdot \Gamma _1)(t,\s,v,\rho ^{-1}(x))
+
(\nabla_{x} \cdot \Gamma _2)(t,\s,v,\rho ^{-1}(x))\\
&=&
-\lambda (h(\rho ^{-1}(x),U(t,\s)))\pp(t,\s,v,\rho ^{-1}(x))\\
&+&
\int_V
\lambda (h(\rho ^{-1}(x),U(t,\s)))T_{h(\rho ^{-1}(x),U(t,\s))}(v,v')\pp(t,\s,v',\rho ^{-1}(x))\,dv'
\eeqn
for all $t,\s,v,x$.
From the definition of $\widetilde \pp$ and the property $h(\rho (\x),\pi u) = h(\x,u)$, which
implies that $h(x,\pi u) = h(\rho ^{-1}(\x),u)$ for all $u$, we conclude that:
\beqn
&&\hspace{-30pt}\frac{\partial \widetilde \pp}{\partial t}(t,\s,v,x)
+
(\nabla_{\s}\cdot \Gamma _1)(t,\s,v,\rho ^{-1}(x))
+
(\nabla_{x} \cdot \Gamma _2)(t,\s,v,\rho ^{-1}(x))\\
&=&
-\lambda (h(x,\pi U(t,\s)))\widetilde \pp(t,\s,v,x)\\
&+&
\int_V
\lambda (h(x,\pi U(t,\s)))T_{h(x,\pi U(t,\s))}(v,v')\widetilde \pp(t,\s,v',x)\,dv'\,.
\eeqn
It will follow that $\widetilde \pp$ is a solution
of~(\ref{eq:Fokker-Planck-Kolmogorov}) with respect to the input field $\pi U$
provided that we show: 
\beqn
(\nabla_{\s}\cdot \Gamma _1)(t,\s,v,\rho ^{-1}(x))
&=&
(\nabla_{\s}\cdot \widetilde \Gamma _1)(t,\s,v,x)\\
(\nabla_{x}\cdot \Gamma _2)(t,\s,v,\rho ^{-1}(x))
&=&
(\nabla_{x}\cdot \widetilde \Gamma _2)(t,\s,v,x),
\eeqn
where
\beqn
\widetilde \Gamma _1(t,\s,v,x) &=& \widetilde \pp(t,\s,v,x)v\\
\widetilde \Gamma _2(t,\s,v,x) &=& \widetilde \pp(t,\s,v,x)f(x,\pi U(t,\s)) \,.
\eeqn
Since $\widetilde \pp(t,\s,v,x) = \pp(t,\s,v,\rho ^{-1}(x))$, the equality for $\nabla_{\s}$
is clear.  We are left to show the equality for $\nabla_{x}$.  We have, fixing
$t,\s,v$ and letting 
$F(x) = f(x,U(t,\s))$,
$C(x) = \pp(t,\s,v,x)$,
$\widetilde F(x) = f(x,\pi U(t,\s))$,
and
$\widetilde C(x) = \widetilde \pp(t,\s,v,x)=C(\rho ^{-1}(x))$:
\beqn
\nabla_{x}\cdot [\widetilde C\widetilde F](x)
&=&
(\partial \widetilde C/\partial x)(x) \widetilde F(x)
\,+\,
\widetilde C(x) (\nabla_{x}\cdot \widetilde F)(x)\\
&=&
(\partial C/ \partial x)(\rho ^{-1}(x)) (\rho ^{-1})_*(x) \widetilde F(x)
\,+\,
C(\rho ^{-1}(x))(\nabla_{x}\cdot \widetilde F)(x)\\
&=&
(\partial C/ \partial x)(\rho ^{-1}(x)) [\rho _*(\rho ^{-1}(x))]^{-1} \widetilde F(x)
\,+\,
C(\rho ^{-1}(x))(\nabla_{x}\cdot \widetilde F)(x)\\
&=&
(\partial C/ \partial x)(\rho ^{-1}(x)) F(\rho ^{-1}(\x))
\,+\,
C(\rho ^{-1}(x))(\nabla_{x}\cdot F)(\rho ^{-1}(\x))\\
&=&
\nabla_{x}\cdot [CF](\rho ^{-1}(x))\,,
\eeqn
where we have used that
$\f(\rho (\x),\pi u) = \rho _*(\x)\f(\x,u)$,
and thus also
$\widetilde F(x)=\rho _*(\rho ^{-1}(\x))F(\rho ^{-1}(\x))$,
as well as the divergence property~(\ref{eq:divergence}).
\epr

In applications, one is often interested in the distribution of positions
irrespective of internal states $x$ and velocities $v$:
\[
Q(t,\s) = \int_\X \int_V \pp(t,\s,v,x) \,d\mu _{\X}(x)\,d\mu _{V}(v)
\]
where $\mu _{\X}$ and $\mu _{V}$ denote appropriate measures on $\X$ and $V$
(and we assume that $\pp$ is integrable).
Take the density corresponding to $\pi U$,
$\widetilde \pp(t,\s,v,x) = \pp(t,\s,v,\rho ^{-1}(x))$, and its marginal
\[
\widetilde Q(t,\s) = \int_\X \int_V \widetilde \pp(t,\s,v,x) \,d\mu _{\X}(x)\,d\mu _{V}(v)
\,.
\]
This is the same as
$
\int_\X \int_V \pp(t,\s,v,x) r(x) \,d\mu _{\X}(x)\,d\mu _{V}(v)
$,
where $r(x)=1/\det \rho _*(x)$.
In the special (but usual in examples) case that $\rho $ is linear, $r$ is a
constant, so $\widetilde Q(t,\s)=rQ(t,\s)$.
It follows that the normalized densities are equal:
\[
\frac{\widetilde Q(t,\s)}{\int \widetilde Q(t,\sigma )\,d\sigma }
\;=\;
\frac{Q(t,\s)}{\int Q(t,\sigma )\,d\sigma }\,.
\]
Alternatively, one could introduce a new measure 
$d\widetilde \mu _{\X}(x) = r(x)\mu _{\X}$, and define $\widetilde Q$ using this new measure, for
all times $t$ and space positions $s$, so that $Q(t,\s) = \widetilde Q(t,\s)$.


\edo

\ee